\documentstyle[psfig,12pt]{report}
\input amssym.def
\input amssym

\begin{document}

\parindent 0mm

\pagestyle{empty}

%*************** PLEASE DO NOT REMOVE THIS STAR LINE **************

\newpage
\begin{flushright}
\end{flushright}
\bigskip
\begin{center} 
\begin{Large} 
\begin{bf}

              Discretization and Continuum Limit\\[0.3cm]
              of Quantum Gravity \\[0.3cm]
              on a Four-Dimensional Space-Time Lattice\\[0.9cm]
            
\end{bf} 
\end{Large}

\begin{large}
      
              Elmar Bittner$^{1,2}$, Wolfhard Janke$^{1}$ and Harald Markum$^{2}$\\[0.9cm]                                              
				  {\em $^{1}$Universit\"at Leipzig, Institut f{\"u}r Theoretische Physik,}\\
				  {\em Augustusplatz 10/11, D-04109 Leipzig, Germany}\\
              {\em $^{2}$Vienna University of Technology, Atominstitut,}\\
              {\em Wiedner Hauptstra{\ss}e 8-10, A-1040 Vienna, Austria}\\

\end{large}
\end{center}

\begin{abstract}
The Regge Calculus is a powerful method to approximate a continuous
manifold by a simplicial lattice, keeping the connectivities of the
underlying lattice fixed and taking the edge lengths as degrees of
freedom. The Discrete Regge Model limits the
choice of the link lengths to a finite number. 
We examine the phase structure of Standard Regge Calculus in four
dimensions and compare our Monte Carlo results with those of the
$Z_2$-Regge Model as well as with another formulation of lattice
gravity derived from group theoretical considerations. Within all
of the three models of quantum gravity we find an extension of the
well-defined phase
to negative gravitational couplings and a new phase transition.
We calculate two-point functions between
geometrical quantities at the corresponding critical point and
estimate the masses of the respective interaction particles.
A main concern in lattice field theories is the existence of a continuum
limit which requires the existence of a continuous phase transition. The
recently conjectured second-order transition of the four-dimensional Regge 
skeleton at negative gravity
coupling could be such a candidate. We examine this regime with Monte Carlo
simulations  and
critically discuss its behavior.
\end{abstract}

\section{Introduction}
The construction of a non-trivial quantum theory of gravitation represents
one of the major open problems in theoretical physics. Approaches
based on perturbative methods failed due to the non-renormalizability of the
underlying theory in dimensions greater than two. One possibility to deal
with the quantization problem consists in discretization of space-time
and quantization via the path integral,
in close analogy to quantum field theory on a flat background geometry.
There are two related schools attempting at a lattice theory of quantum
gravity, dynamical triangulation and Regge quantum gravity. We concentrate
on the latter approach and refer to \cite{cat} for an excellent review of
the former framework. Regge calculus is the only discretization scheme
reproducing the Bianchi identities of classical general relativity and
is formulated exclusively in terms of the edge lengths of the lattice,
independent of any coordinate system. Taking advantage of the freedom in
how to choose the lattice structure as well as the action,
we investigate in this paper different formulations, 
all of them based on the original work by Regge \cite{reg}, namely 
Standard Regge Calculus (SRC) \cite{ber1,ham1}, a Group Theoretical Model 
(GTM) \cite{cas}, and the $Z_2$-Regge Model ($Z_2$RM) \cite{juri,mark}. 
Different to ordinary lattice gauge theory, quantum fluctuations are
represented by fluctuations in the edge lengths.
In particular the Monte Carlo method is a convenient tool for evaluating
the discrete Euclidean functional integral in a
non-perturbative way. It was shown that within a certain range of the
gravitational coupling the path integral converges to make up a well-defined
phase of lattice quantum gravity. We present an extension of the phase 
diagram of the models in four dimensions mentioned above.
Due to the fact that a Wick rotation from the Lorentzian to 
the Euclidean sector of quantum gravity is not feasible, in general the 
sign of the action in the path integral is not fixed {\em a priori}.
Therefore we investigate the continuation of the well-defined phase into
the region of negative gravitational coupling. It turns out to be
terminated by a pronounced transition-like cross-over to an ill-defined phase suggesting to look for a
continuum limit. In particular the $Z_2$RM permits accurate investigations of
the negative coupling region.
Further, we calculate two-point functions
to probe the existence of massless quanta of the gravitational field. A
candidate for a realistic quantum theory of gravity should reproduce the
expected long-range interaction behavior observed in nature.

\section{Lattice quantum gravity}
Any smooth $d$-manifold can be approximated by appropriately gluing together
pieces of flat space, called $d$-simplices, ending up with a simplicial
lattice. We take the edge lengths as the dynamical degrees of freedom and
leave the triangulation of the lattice fixed. Adopting the Euclidean
path-integral approach we may write down the partition function
\begin{equation} \label{Z}
Z=\int D[q]e^{-I(q)} ~,
\end{equation}
with the gravitational action $I$ introduced
in more detail below. The functional integration extends over the squared
edge lengths $q_l$ of the links $l$ of the simplicial lattice.

One of the problems with (\ref{Z}) is the ambiguity in performing the
link-lengths integration. Commonly the measure is written as
\begin{equation} \label{Dq}
D[q]=\prod_l dq_lq_l^{\sigma-1}{\cal F}(q) ~,
\end{equation}
with ${\cal F}$ a function of the quadratic edge lengths being equal
to one if the Euclidean triangle inequalities are fulfilled and zero
otherwise. The question remains whether such a local measure is
sufficient at all and how the power $\sigma$ might be chosen. In an 
effort to clarify the role of the measure the conventional definition 
of diffeomorphisms has been employed in the two-dimensional case, 
assuming that a piecewise linear space, i.e.~a Regge surface, is exactly
invariant under the action of the full diffeomorphism group \cite{men}. 
After a conformal gauge fixing was performed in the continuum formalism
with the DeWitt measure, 
it was shown that the evaluation of the non-local Faddeev-Popov determinant
by using such a Regge regularization leads to the usual Liouville field 
theory results in the continuum limit. All that is based on a description
of piecewise linear manifolds with deficit angles, not edge lengths, and
is mostly taken as an argument that the correct measure of Standard Regge
Calculus has to be non-local. However, to our knowledge it is not obvious
that this argument carries over to a discretized Lagrangian, which is 
formulated in terms of fluctuating edge lengths, obeying triangle 
inequalities, and which is not invariant under the diffeomorphism group 
due to the presence of curvature defects: different assignments of edge 
lengths correspond to different physical geometries, and as a consequence 
there are no gauge degrees of freedom in Standard Regge Calculus, apart 
from special geometries (like flat space) \cite{hartle,holm}.
Furthermore, the generalization of this procedure
to higher dimensions and its numerical implementation are technically
demanding and some of the proposed non-local measures do not
agree with their continuum counterparts in the weak field limit, which
is a necessary condition for an acceptable discrete measure \cite{hw97}.
But this property is fulfilled for the standard simplicial measure 
(\ref{Dq}).

Working in Euclidean space, i.e.~with positive definite metric,
the conformal mode renders the four-dimensional continuum Einstein-Hilbert action
unbounded from below. This unpleasant feature persists in the
discretized Regge-Einstein action but need not necessarily lead to an
ill-defined path integral \cite{ber1}. Indeed, numerical simulations
reveal the existence of a well-defined phase with finite expectation
values within a certain range of the bare gravitational coupling.
A lattice action for gravitation is given by
\begin{equation} \label{Iq}
I(q)=-\beta\sum\limits_tR_t(A_t,\delta_t)+\lambda\sum\limits_sV_s ~,
\end{equation}
where the first sum runs over all triangles $t$ with areas $A_t(q)$ and 
the corresponding 
deficit angles $\delta_t(q)$ yield the curvature elements $R_t$. The 
second term extends over the volumes $V_s(q)$ of all four-simplices $s$ 
and allows together with the cosmological constant $\lambda$ to set an 
overall scale in the action. Next we define our three different models
for subsequent numerical treatment.

\subsection{Standard Regge Calculus (SRC)}
Here we employ the Regge-Einstein action with $R_t=A_t\delta_t$, set
the cosmological constant $\lambda=1$ and choose the gravitational
measure to be uniform, $\sigma=1$ \cite{ber1,ham1}. In the classical 
continuum limit the Regge action $2\sum_t R_t$ is equivalent to the 
Einstein-Hilbert action $\int d^4x\sqrt{g}R$ if the so-called fatness 
$\phi_s$ of a 4-simplex obeys \cite{chee}
\begin{equation}
\phi_s\sim\frac{V_s^2}{\mbox{max}_{l\in s}(q_l^4)}\ge f=\mbox{const}>0 ~.
\end{equation}
A lower limit $f=10^{-4}$ restricts the configuration space and thus
facilitates numerical simulations \cite{we}. 

\subsection{Group Theoretical Model (GTM)}
Constructing the dual of a simplicial lattice, Poincar{\'e} transformations
can be assigned to its links to yield an action in which the {\em sin}
of the deficit angle enters, \hbox{$R_t=A_t\sin\delta_t$}, and that reduces 
to the Regge action in the small curvature limit \cite{cas}. While in the
classical continuum limit the deficit angles become small, this is in 
general not the case in the path-integral quantization on a finite lattice,
where one has to sum over all possible discrete configurations. 
{\em A priori} it is hence not clear whether both SRC and the GTM result in the
same quantum continuum theory. For comparison we again use
the uniform measure with $\sigma=1$, set the cosmological constant 
\hbox{$\lambda=1$}, and choose a lower limit on the fatness, $f=10^{-4}$. 

\subsection{$Z_2$-Regge Model ($Z_2$RM)}
This model was invented in an attempt to reformulate SRC as the partition
function of a spin system \cite{juri,mark}. It is defined by restricting
the squared link lengths to take on only two values
\begin{equation} \label{q_l}
q_l=b_l(1+\epsilon\sigma_l) ~,\quad \sigma_l\in Z_2 ~.
\end{equation}
By setting $b_l=1,2,3,4$ for edges, face diagonals, body diagonals and the 
hyperbody diagonal of a hypercube, respectively, the link lengths are 
allowed to fluctuate around their flat-space values. The Euclidean triangle 
inequalities are automatically fulfilled as long as 
$\epsilon\le\epsilon_{\rm max}\in I\!\! R_+$ and therefore ${\cal F}=1$ in any 
case. Consequently, the uniform measure in the quantum-gravity 
path-integral becomes identical to unity for all possible link configurations. 
The action can be rewritten in terms of complicated local ``spin-spin'' 
interactions. Nevertheless,  numerical simulations of the $Z_2$RM
 become extremely efficient by
implementing look-up tables and a heat-bath algorithm. Computations
have been performed with the parameter $\epsilon=0.0875$ and the 
cosmological constant $\lambda=0$ because (\ref{q_l}) already fixes
the average lattice volume.

\newpage
\section{Phase structure}
Monte Carlo simulations with at least 50k sweeps for each value of the
coupling $\beta$ have been performed on regularly triangulated hypercubic
lattices with toroidal topology and $N_0=4^4$ vertices. We measured
expectation values of the average link length 
$\langle q\rangle=\langle N_1^{-1}\sum_lq_l\rangle$ ($N_1$ denotes the
total number of links) and the average curvature
\begin{equation}
{\langle R\rangle}=\left\langle\frac{2\sum_t R_t}
                   {\sum_s V_s} q\right\rangle ~,
\end{equation}
within the models described above. 

\begin{figure}[h]
\centerline{\psfig{figure=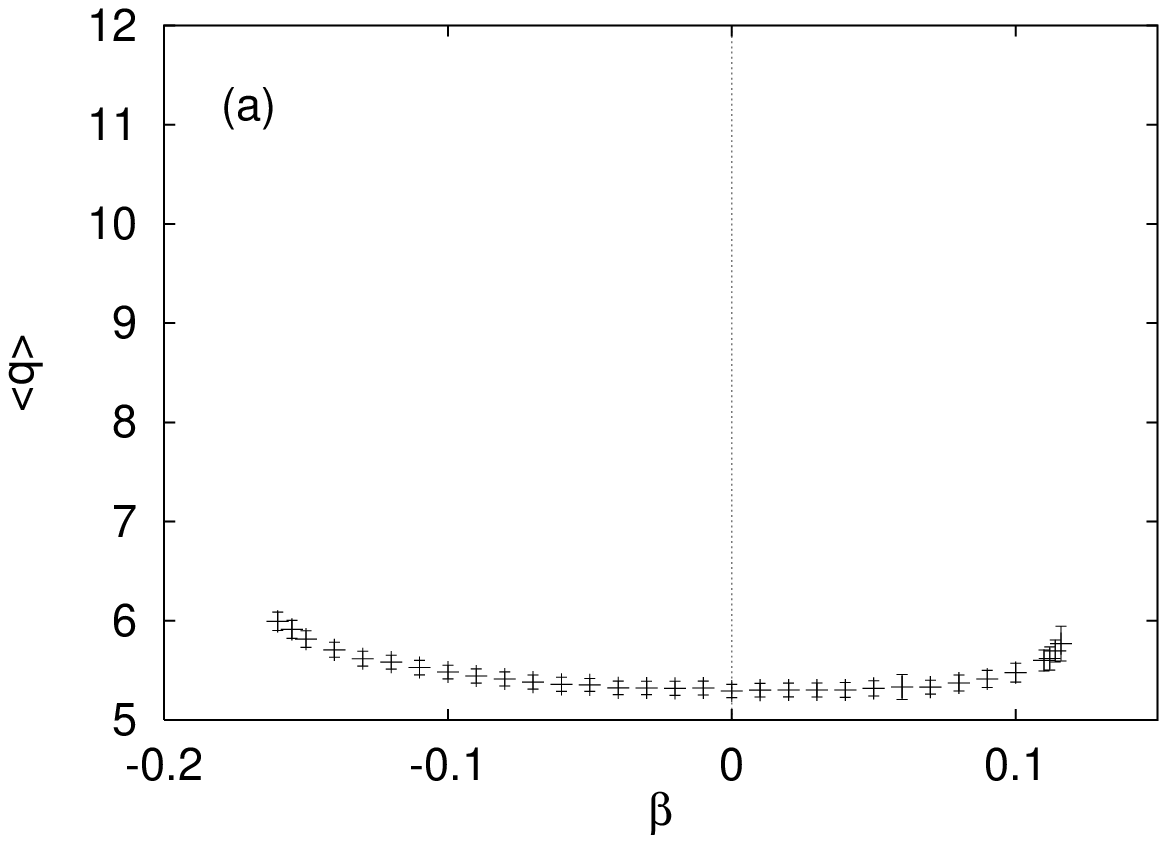,height=6cm,width=8cm}}
\centerline{\psfig{figure=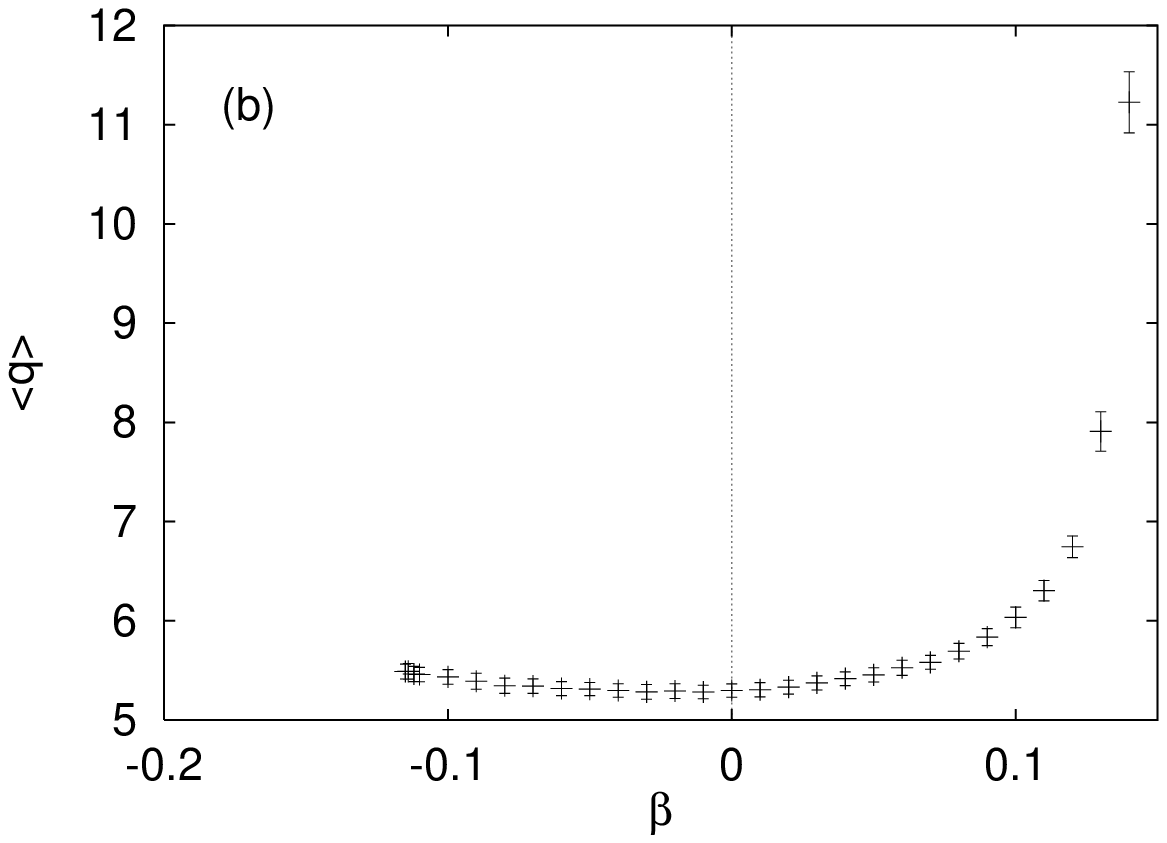,height=6cm,width=8cm}}
\centerline{\psfig{figure=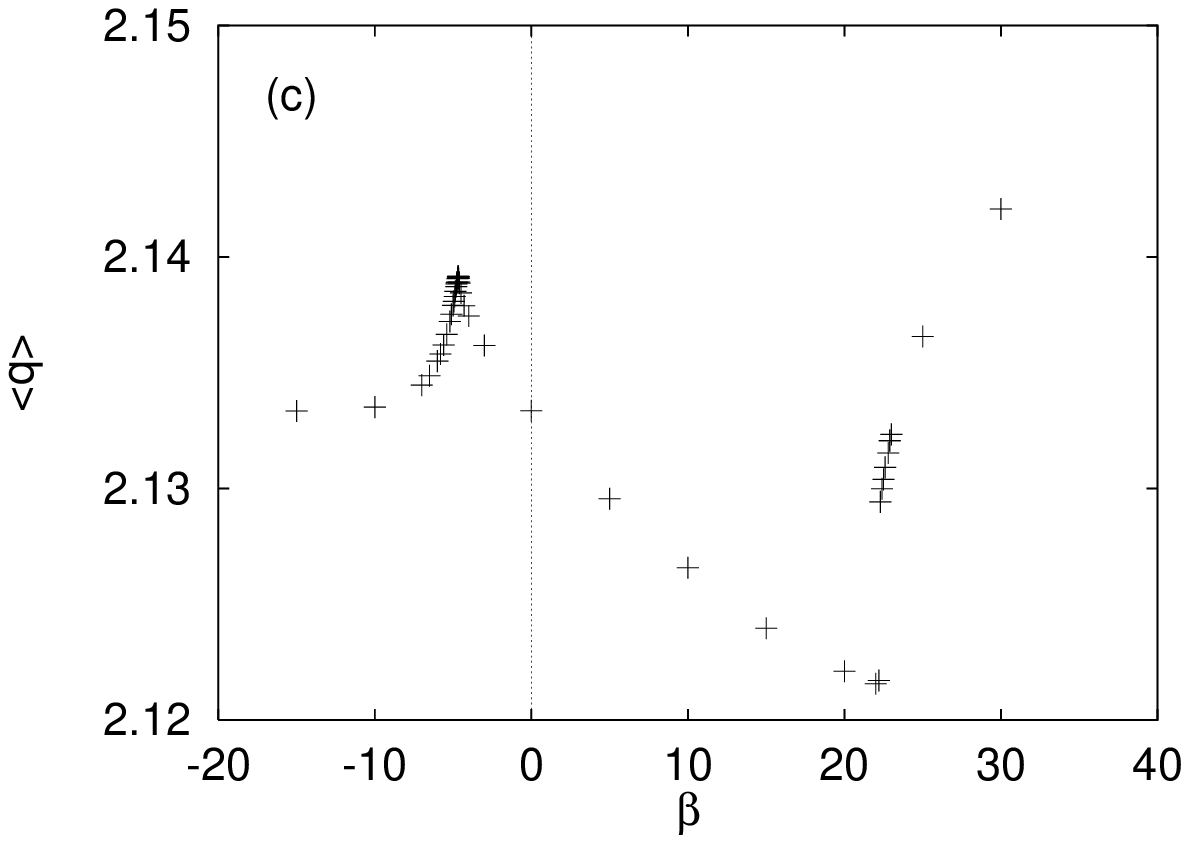,height=6cm,width=8cm}}
\vspace{3mm}
\caption{Expectation values of the average link lengths as a function of 
the gravitational coupling for (a) SRC, (b) GTM, and (c) $Z_2$RM.}
\label{qfigs}
\end{figure}

\begin{figure}[h]
\centerline{\psfig{figure=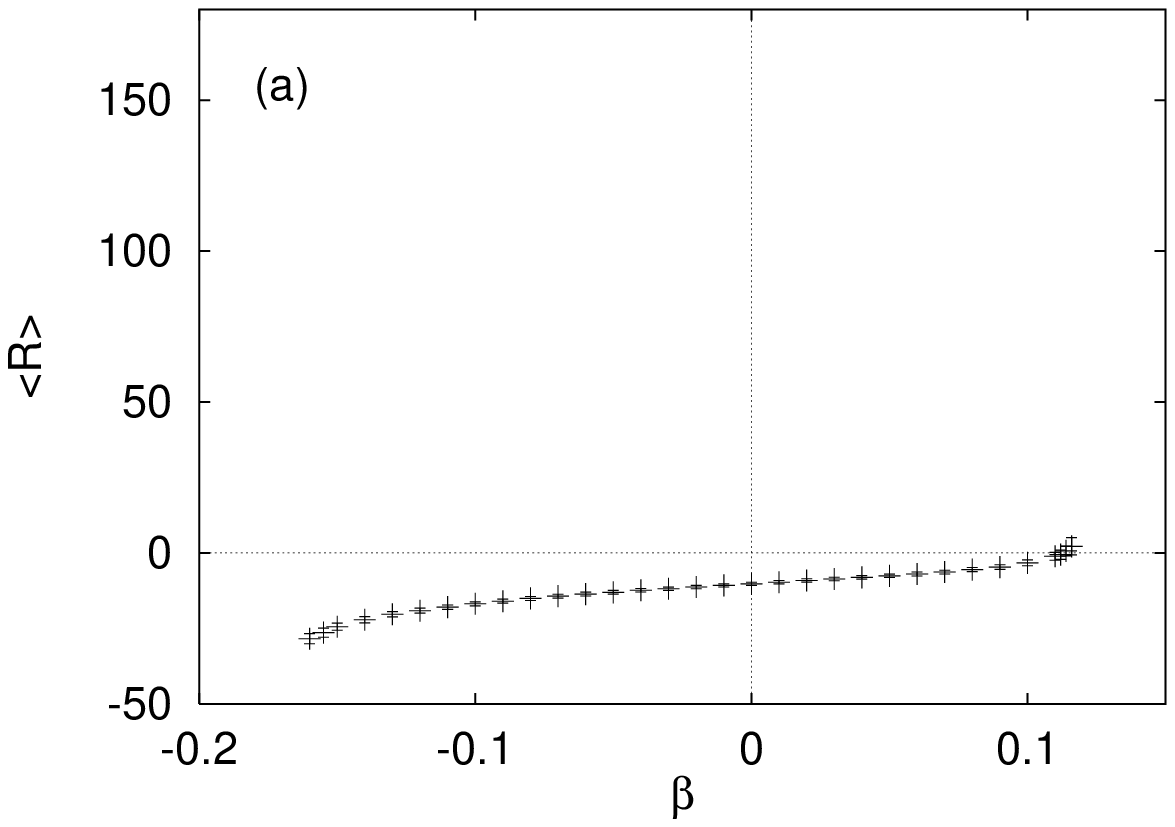,height=6cm,width=8cm}}
\centerline{\psfig{figure=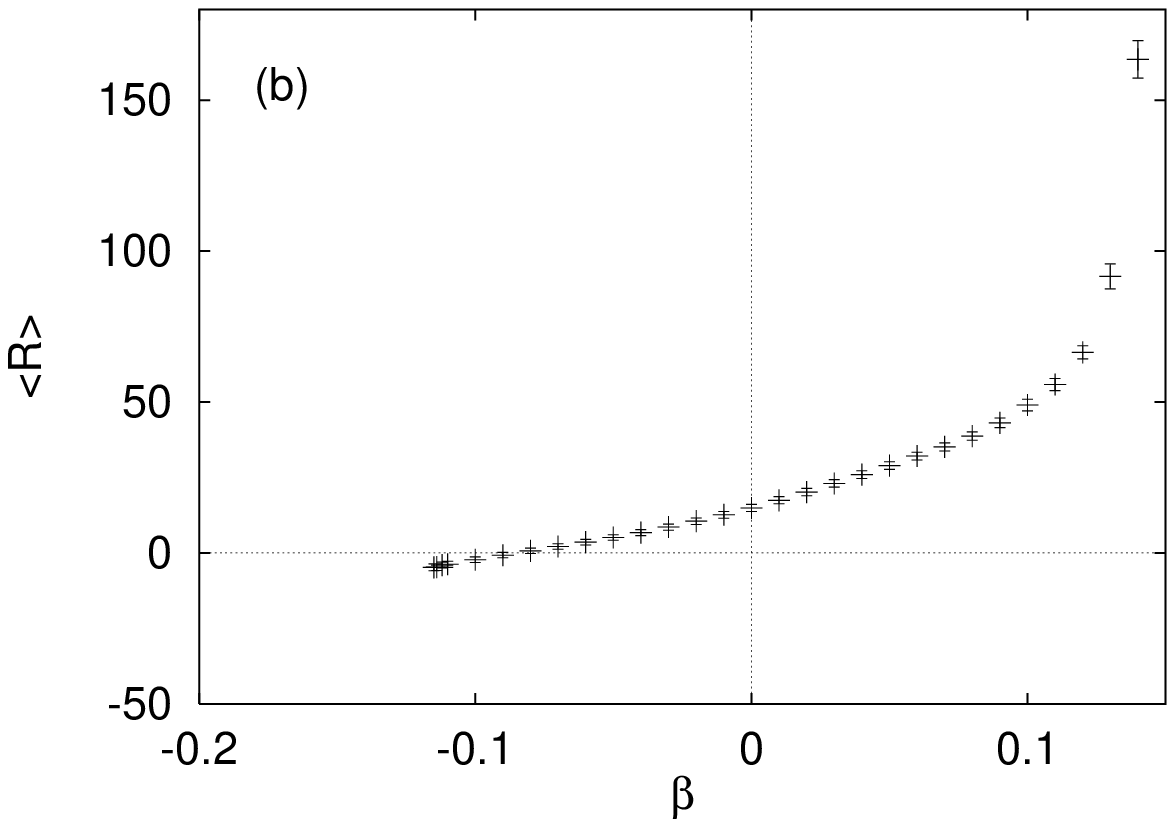,height=6cm,width=8cm}}
\centerline{\psfig{figure=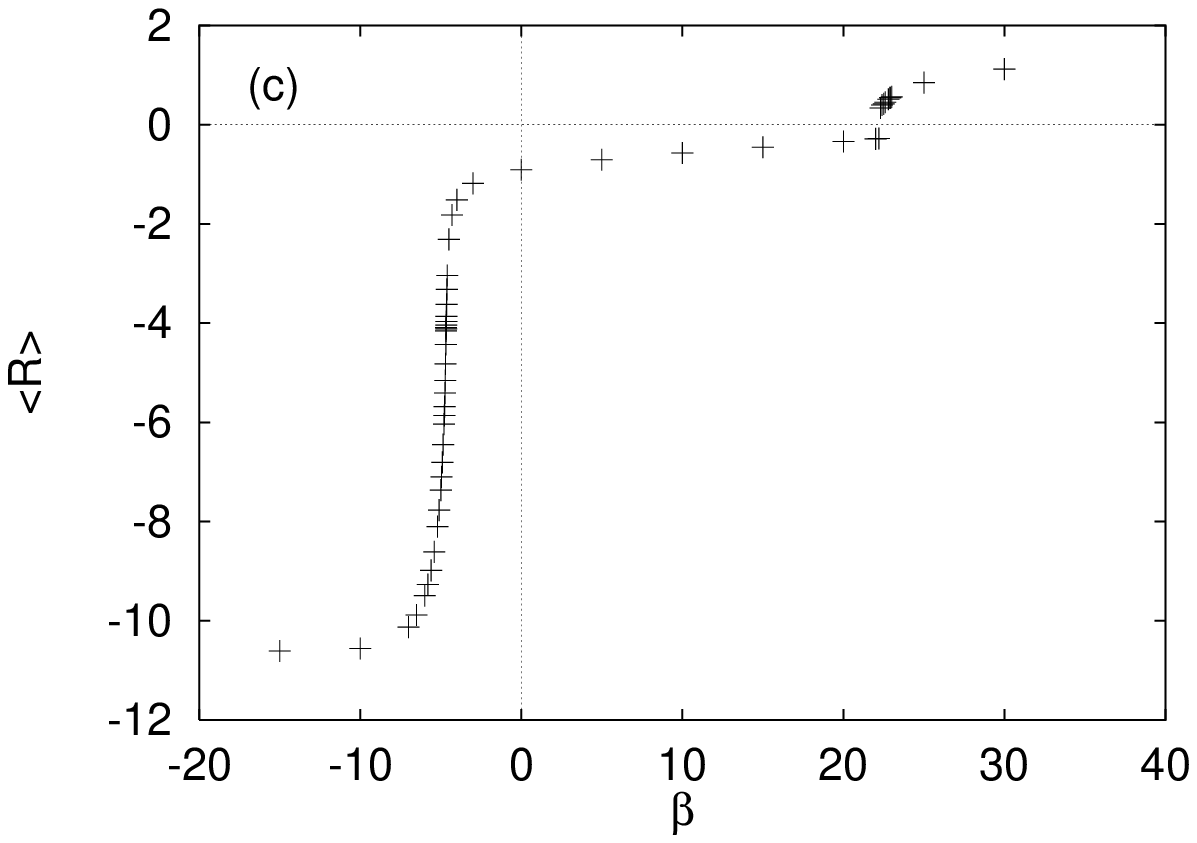,height=6cm,width=8cm}}
\vspace{3mm}
\caption{Expectation values of the average curvature as a function of the
gravitational coupling for (a) SRC, (b) GTM, and (c) $Z_2$RM. All
models exhibit a related phase structure and could lie in the same
universality class.}
\label{Rfigs}
\end{figure}

Let us first concentrate on SRC, cf. figures \ref{qfigs}a and \ref{Rfigs}a. 
The expectation values of the average link length change only slightly,
increase near $\beta_c^-\approx-0.16$ and $\beta_c^+\approx 0.116$,
and become arbitrarily large for $\beta<\beta_c^-$ and $\beta>\beta_c^+$ 
indicating the emergence of an ill-defined phase with spike-like structures. 
The expectation value of the average curvature is negative in the 
well-defined phase, but $|\langle R\rangle|$ becomes very large in the 
ill-defined phase where simplices collapse into degenerate configurations
with long links and small volumes.
The coupling of SU(2)-gauge theory to SRC was examined in \cite{su2} and
turns out to have little influence on the gravitational phase diagram.

The GTM behaves very similar, see figures \ref{qfigs}b
and \ref{Rfigs}b. Again expectation values are finite and well-defined
only in a certain coupling region between $\beta_c^-\approx-0.115$ and 
$\beta_c^+\approx 0.14$. In contrast to the Regge model the curvature 
$\langle R\rangle$ is mostly positive in this coupling interval.
However, as long as we deal with non-renormalized entities it is not clear
whether the observed differences in the phase structure are physically
significant. Although SRC and GTM have the same classical continuum limit,
it is necessary to demonstrate that they coincide in their critical
properties in order to have the same quantum continuum limit.

\begin{figure}[h]
\centerline{\psfig{figure=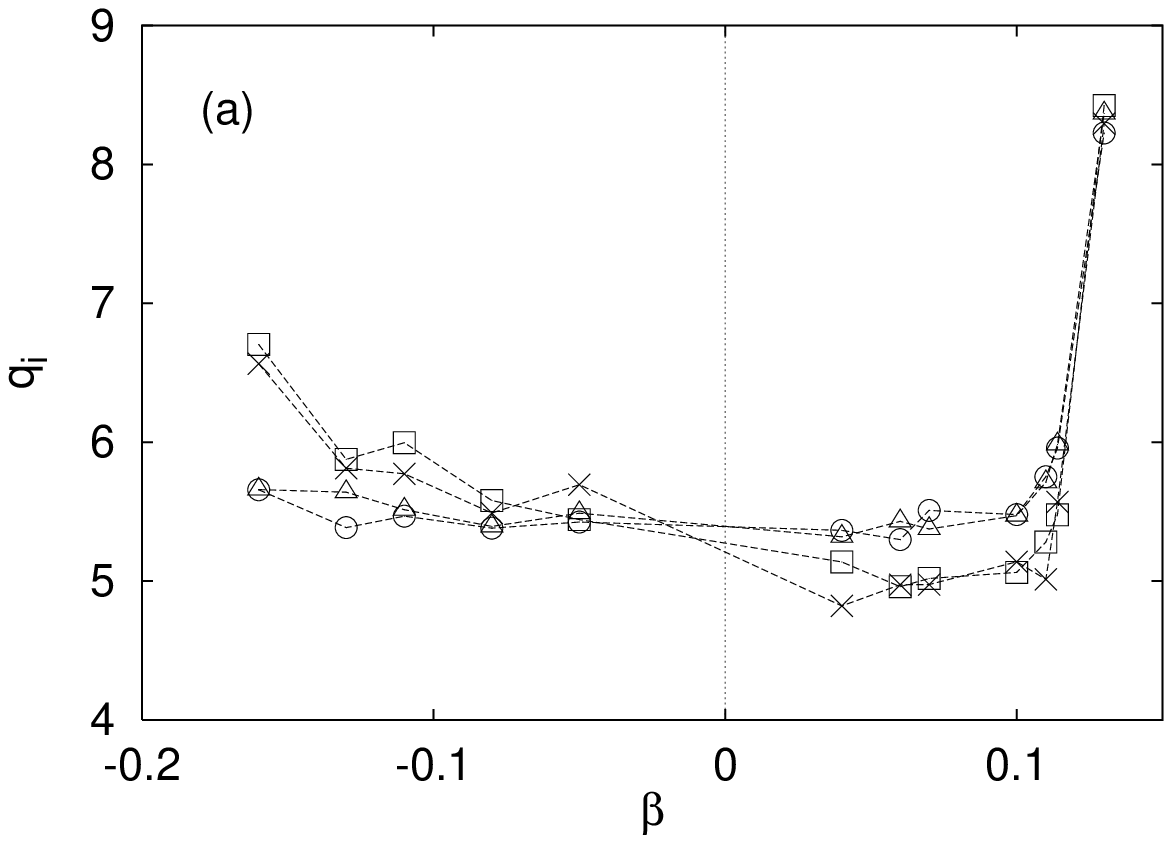,height=6cm,width=8cm}}
\centerline{\psfig{figure=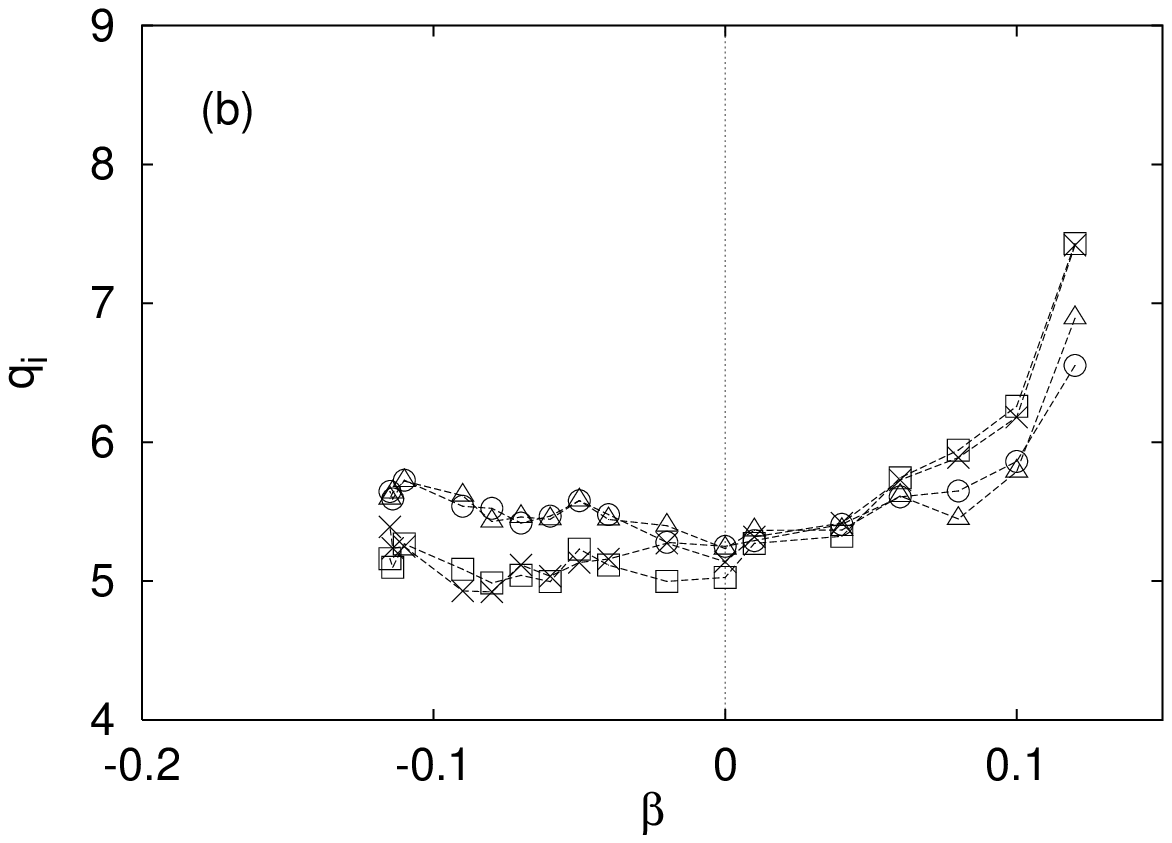,height=6cm,width=8cm}}
\centerline{\psfig{figure=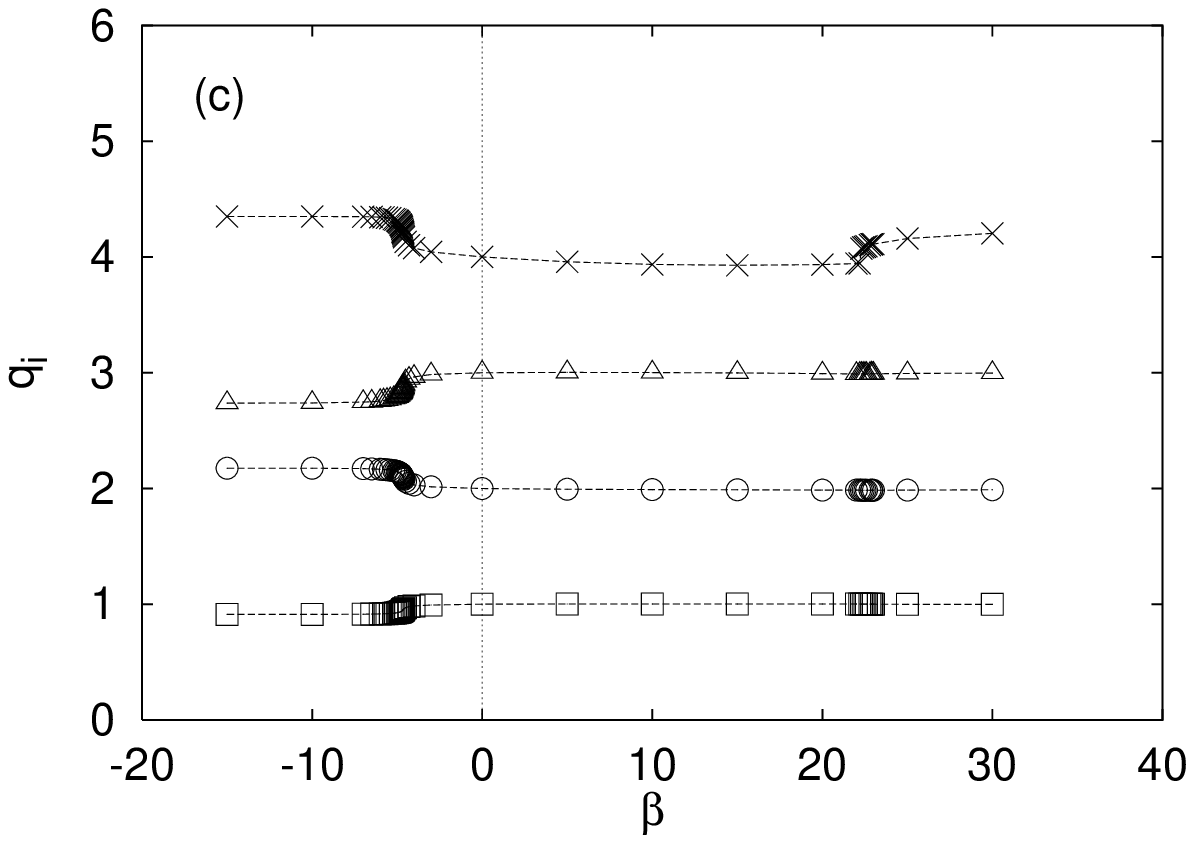,height=6cm,width=8cm}}
\vspace{3mm}
\caption{Lattice average of the squared link length of type $i$ as a
function of the gravitational coupling for (a) SRC, (b) GTM, and
(c) $Z_2$RM. The symbol $\Box$ denotes the edges, $\circ$ the face diagonals,
$\triangle$ the body diagonals, and $\times$ the hyperbody diagonal
of a triangulated hypercube. The dashed lines are to guide the eyes.}
\label{qlfigs}
\end{figure}

The $Z_2$RM is always well-defined because (\ref{q_l}) limits the link
lengths {\em a priori}, cf. figures \ref{qfigs}c and \ref{Rfigs}c. The 
lattice freezes, forming characteristic configurations where the systems
with continuously varying edge lengths become ill-defined. Such a model
is of course better suited for investigations of the phase boundaries which 
here occur at $\beta_c^-\approx-4.665$ and $\beta_c^+\approx 22.3$. While
the transition at $\beta_c^+$ is quite clearly of first order, the
behavior at $\beta_c^-$ is more subtle and will be discussed in some detail below in 
Section~\ref{sec_phase}.
Different to SRC and the GTM the four kinds of links in the $Z_2$RM are
not equivalent by definition (\ref{q_l}). This is reflected in the
corresponding average link lengths $q$ in a single configuration, see 
figure \ref{qlfigs}. Thus we can examine their respective influence on 
the phase transitions. For example the hyperbody diagonal shows a 
considerable discontinuity at $\beta_c^+$ and exactly this diagonal was
identified from weak-field calculations of Regge gravity to belong to
the (five) spurious metric degrees of freedom \cite{rowi}. It would be
interesting to clarify whether physical and spurious degrees of freedom
decouple as predicted from weak-field Regge theory. A simulation of
the $Z_2$RM with the hyperbody diagonal fixed results in 
unchanged order of the transition but in slightly shifted 
values of the critical couplings. An intriguing question remains: to 
what extent are the (ten) physical degrees of freedom responsible for 
(the order of) the phase transitions?

A related issue was investigated in \cite{wb} using the SRC on general,
non-regular triangulations of the four-torus. Also there, even with 
additional higher-order terms in the action, the phase transition is
influenced by the local coordination numbers.

\section{Correlation functions}
One important feature of a physically relevant theory of quantum gravity
should be the existence of a massless graviton. The gravitational
weak-field propagator can be cast into a spin-zero and a spin-two part
\cite{ham2}, with the latter written in terms of connected curvature-curvature
correlations, 
\begin{equation} \label{GR}
G_R(d)=\langle\sum_{t\supset v_0}R_t\sum_{t'\supset v_d}R_{t'}
\rangle_c ~.
\end{equation}
The local operators should be measured at two vertices
$v_0$ and $v_d$ separated by the geodesic distance $d$, which we take
to be equal to the index distance along the main axes of the skeleton.
This seems a reasonable approximation in the well-defined phase with its
small average curvature. In general one expects for (\ref{GR}) at large
distances the functional form
\begin{equation} \label{G}
G_R(d)\sim\frac{e^{-md}}{d^a} ~.
\end{equation}
A power law with $a=2$ and a vanishing effective mass $m=0$ would hint
at \mbox{Newtonian} gravity with massless gravitons. Indeed this was found 
in \cite{ham2} for SRC with higher-order curvature terms at some positive
critical coupling. In contrast to that, fast decaying exponentials were
reported for (\ref{GR}) and also for volume correlations in the case of
SRC at $\beta_c^+$ \cite{lois}. 

\begin{figure}[h]
\centerline{\psfig{figure=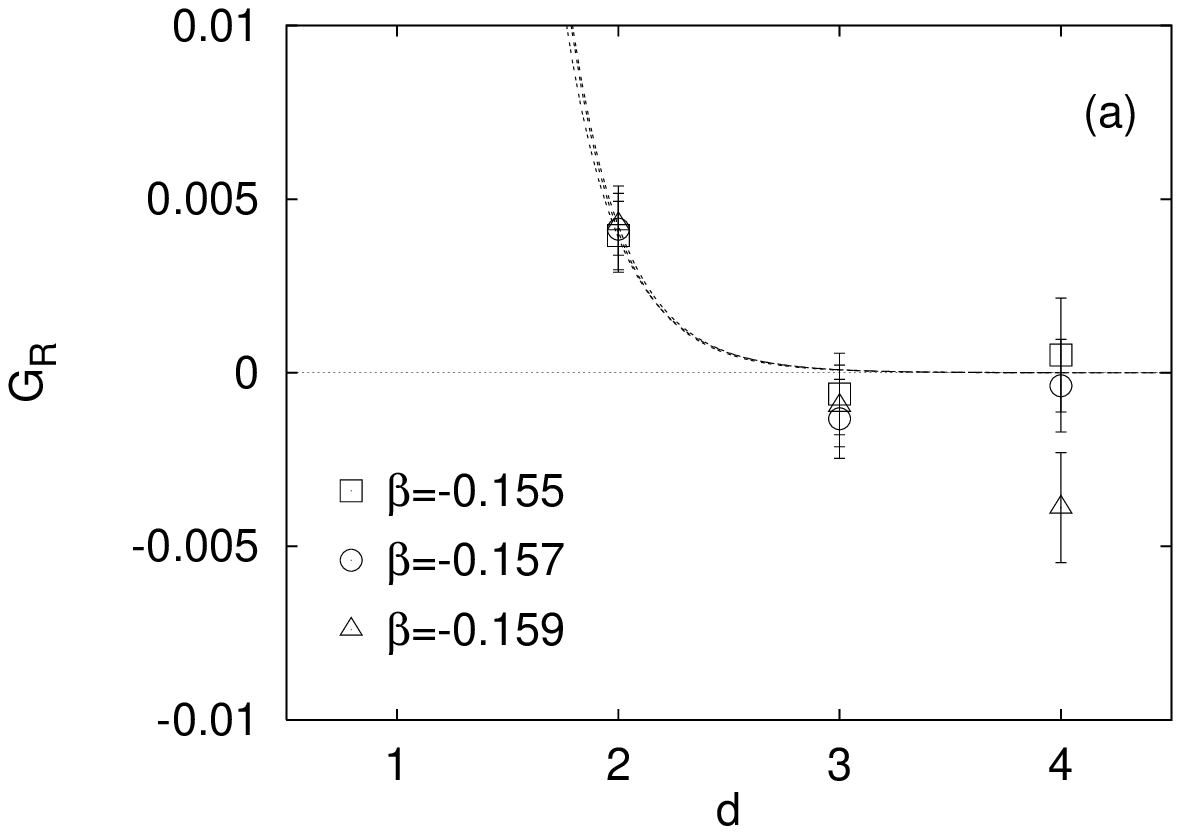,height=6cm,width=8cm}}
\centerline{\psfig{figure=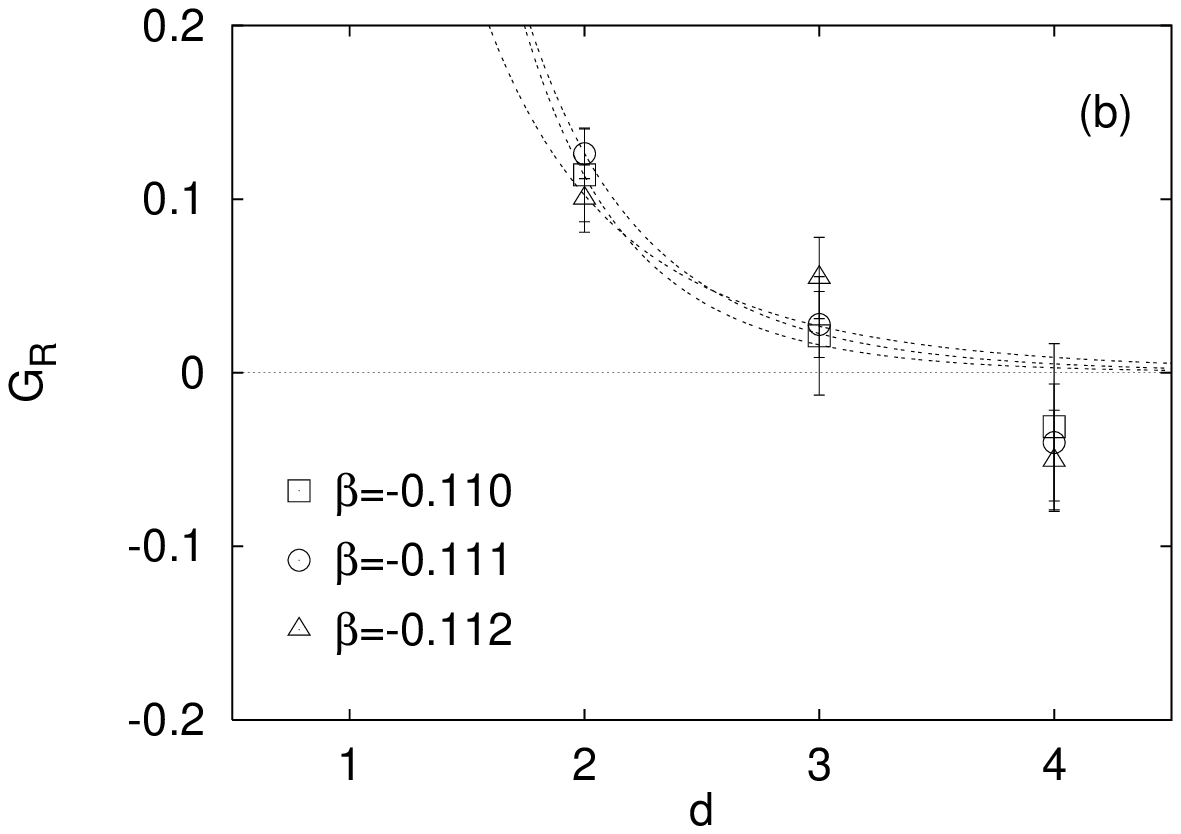,height=6cm,width=8cm}}
\centerline{\psfig{figure=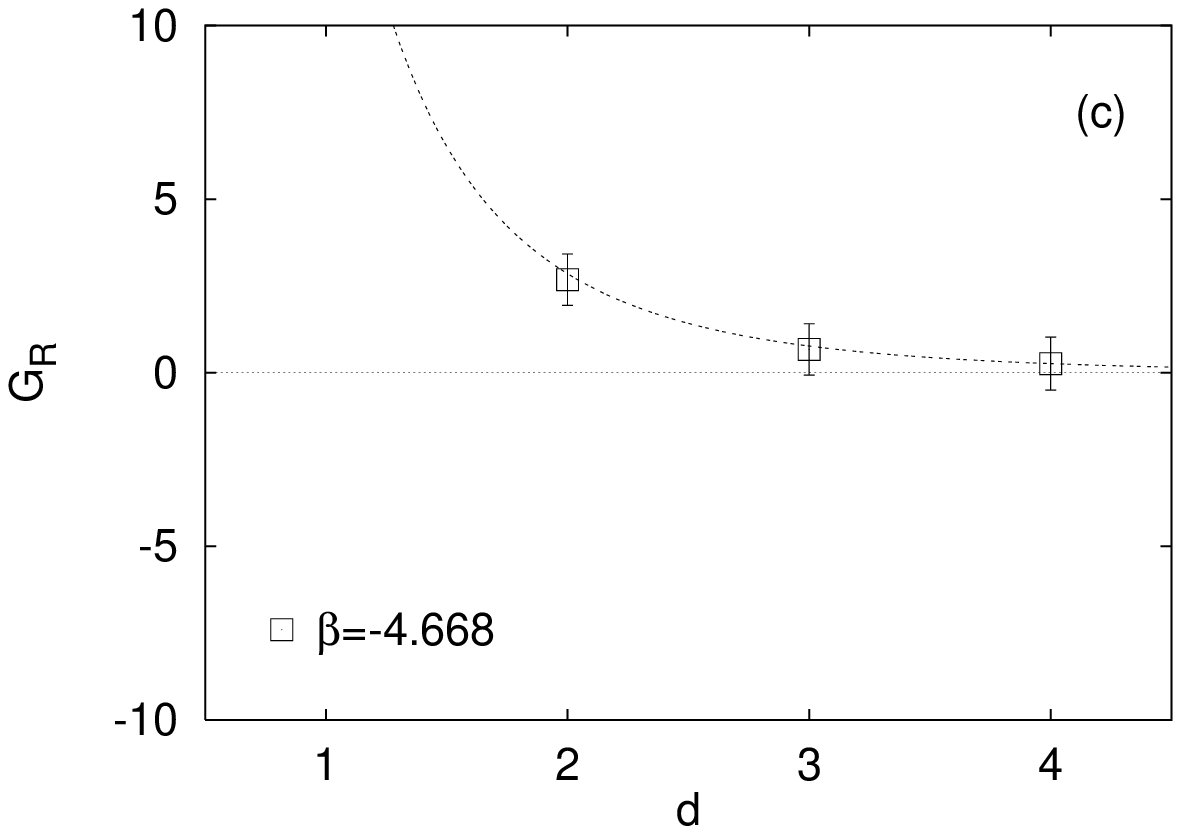,height=6cm,width=8cm}}
\vspace{3mm}
\caption{Curvature correlation functions for (a) SRC, (b) GTM, and
(c) $Z_2$RM in the vicinity of $\beta_c^-$.
The curves show fits to the Yukawa ansatz (\ref{G}) with fixed parameter $a=2$.
The resulting masses $m$ are compiled in table 1.}
\label{figr}
\end{figure}

We now compare SRC and GTM (on $4^3\times 8$ lattices) with $Z_2$RM
(on $8^4$ lattice) \cite{letz}.  The gravitational couplings were chosen 
close to the negative critical coupling
$\beta_c^-$ where there is a chance for a continuous phase transition
and thus a continuum limit. Figure \ref{figr} displays our Monte Carlo data
for the two-point functions (\ref{GR}). We took only distances $d\ge 2$
into account because $G_R(d=1)$ is plagued by lattice artifacts due to
contact terms. In order to test whether (\ref{GR}) obeys a power law we
fixed $a=2$ and fitted the effective masses $m$, cf. table \ref{tab2}.
For SRC $m$ remains rather constant towards the critical coupling whereas
in the GTM the mass decreases for $\beta\to \beta_c^-$. However for all
fits the uncertainties in the mass parameters are large allowing even
for $m=0$. For both the $Z_2$RM and the GTM, power-law fits to (\ref{G})
are compatible with vanishing effective mass $m$. Also for SRC an algebraic decay
is compatible but, due to the large uncertainties, obviously no
definite conclusion can be drawn here. 

\begin{table}[hb]
\caption{\label{tab2} Effective masses $m$ of the curvature two-point
 function $G_R(d)$ for gravitational couplings close to $\beta_c^-$.}
\begin{itemize}
\item[]\begin{tabular}{@{}l@{\extracolsep{\fill}}lllllllll}\hline\hline
 &\multicolumn{3}{c}{SRC} &~& \multicolumn{3}{c}{GTM} &~& 
 \multicolumn{1}{l}{$Z_2$RM}\\\hline
$\beta~~$~ & $-0.155$ & $-0.157$ & $-0.159$ &~& $-0.110$
& $-0.111$ & $-0.112$ &~& $-4.668$ \\
$m$ & ~~$\,2(9)$ & ~~$\,3(24)$ & ~~$\,3(17)$ &~& ~~$\,1.0(21)$ & ~~$\,0.9(8)$
& ~~$\,0.5(7)$ &~& ~~$\,0.5(9)$ \\\hline\hline
\end{tabular}
\end{itemize}
\end{table}

\newpage
\section{Continuum limit at negative gravitational coupling}\label{sec_phase}
\begin{figure}
\centerline{\psfig{figure=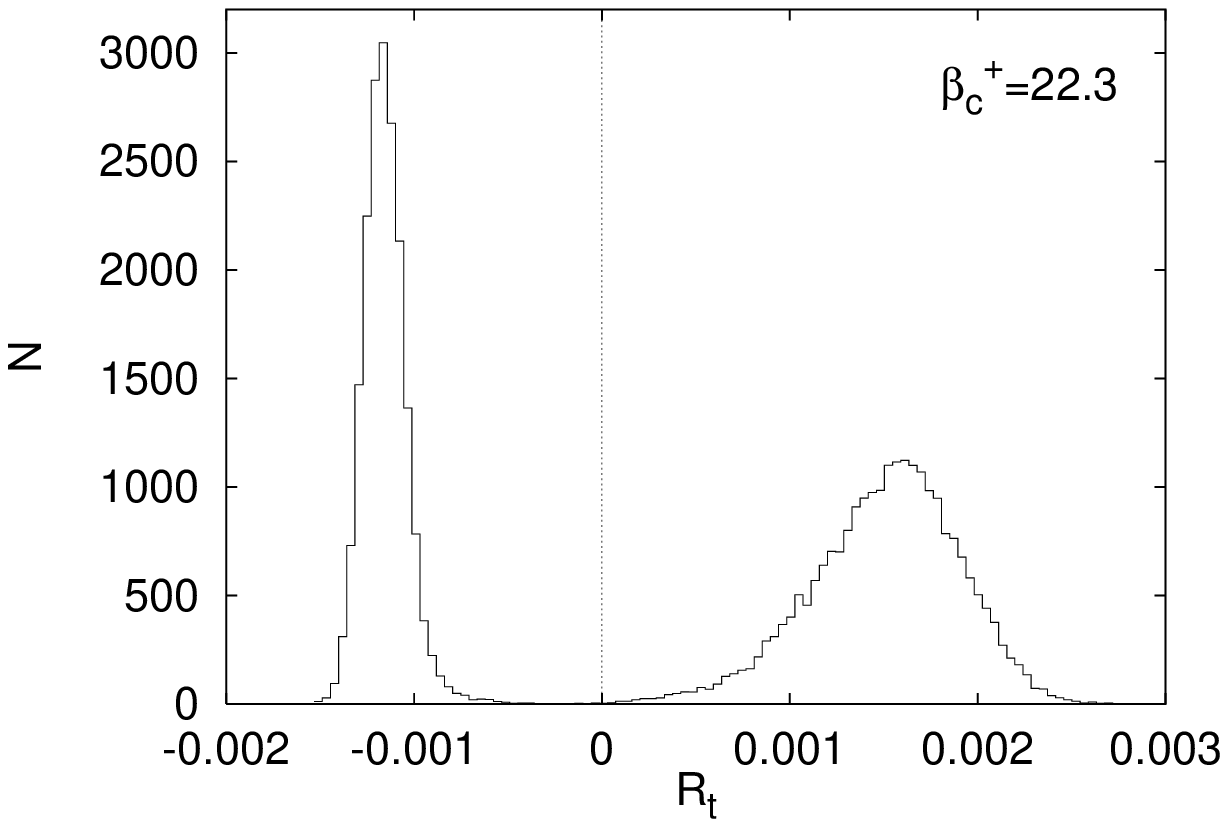,height=6cm,width=8cm}}
\centerline{\psfig{figure=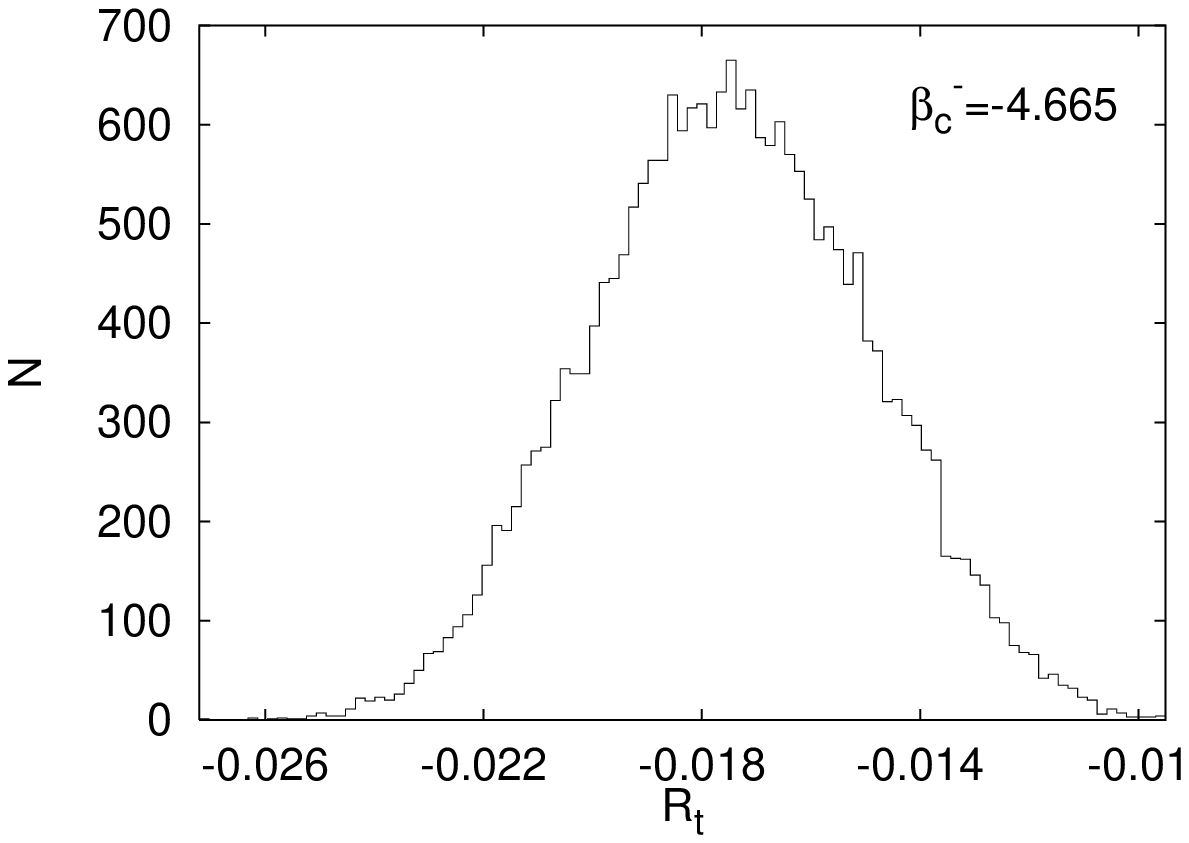,height=6cm,width=8cm}}
\vspace{3mm}
\caption{Histograms of the Regge action $R_t=A_t\delta_t$ from simulations
of the $Z_2$-Regge Model. The distributions indicate a first-order transition
around \hbox{$\beta_c^+=22.3$} (upper plot) and a continuous transition around
$\beta_c^-=-4.665$ (lower plot).}
\label{his}
\end{figure}

The Discrete Regge model $Z_2$RM -- like full Regge theory -- exhibits two phase 
transitions \cite{mel}. One is located
at a negative value and  the other one at a positive value of the bare 
gravitational coupling.
Although earlier work concentrated on the latter transition~\cite{hamber}, there is no reason 
for favoring
a positive value of $\beta$ over a negative one due to the Wick rotation problem 
mentioned in the Introduction.
In the vicinity of the transition
at positive $\beta$, histograms, e.g.~of $A_t\delta_t$ \cite{mel}, clearly show a two-peak
structure, see figure~\ref{his}. The two phases coexist and tunnelling from one phase to the other
and back takes place. The system also exhibits a hysteresis; the transition occurs
at a larger value of $\beta$ if the simulation is started from a configuration
in the well-defined phase than it does if the calculation
is started from a ``frozen'' configuration. Given all these pieces of evidence, 
we conjecture that the transition at $\beta > 0$ is of first order.

\begin{figure}[t]
\centerline{\psfig{figure=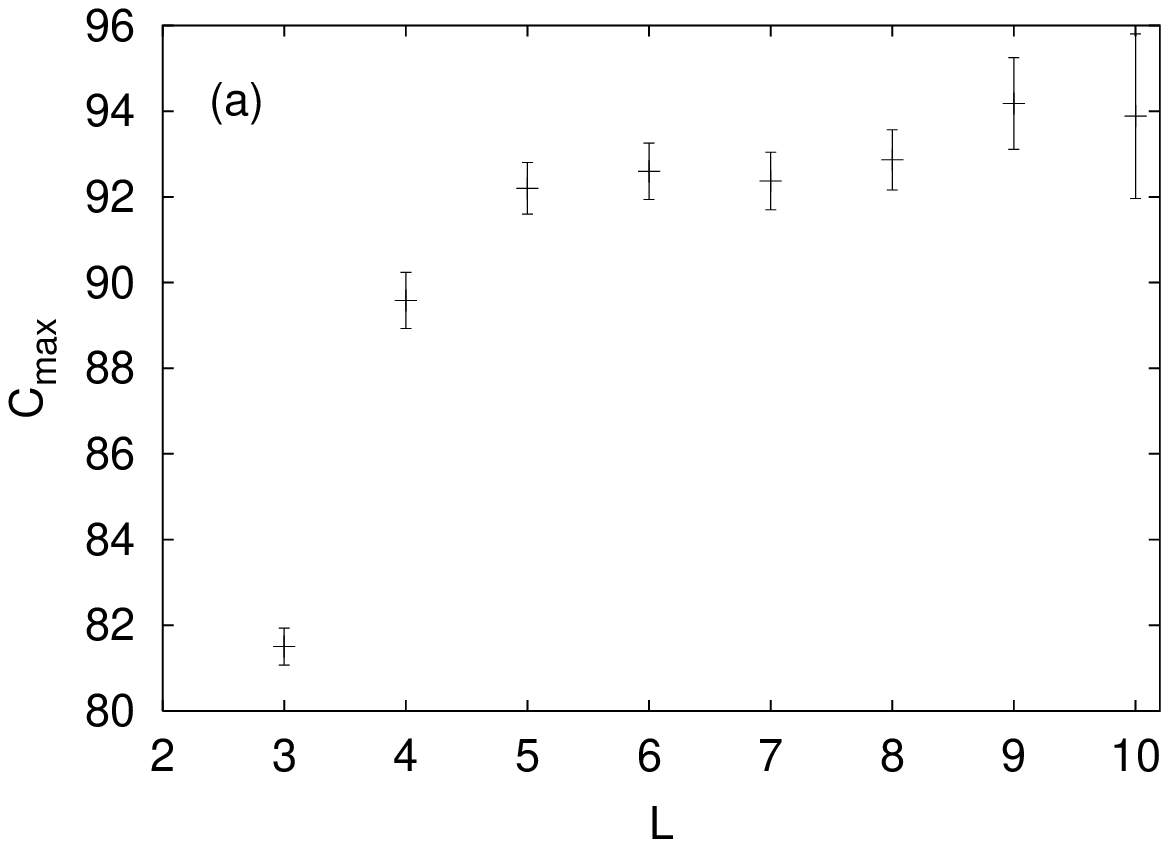,width=8cm,height=6cm}}
\centerline{\psfig{figure=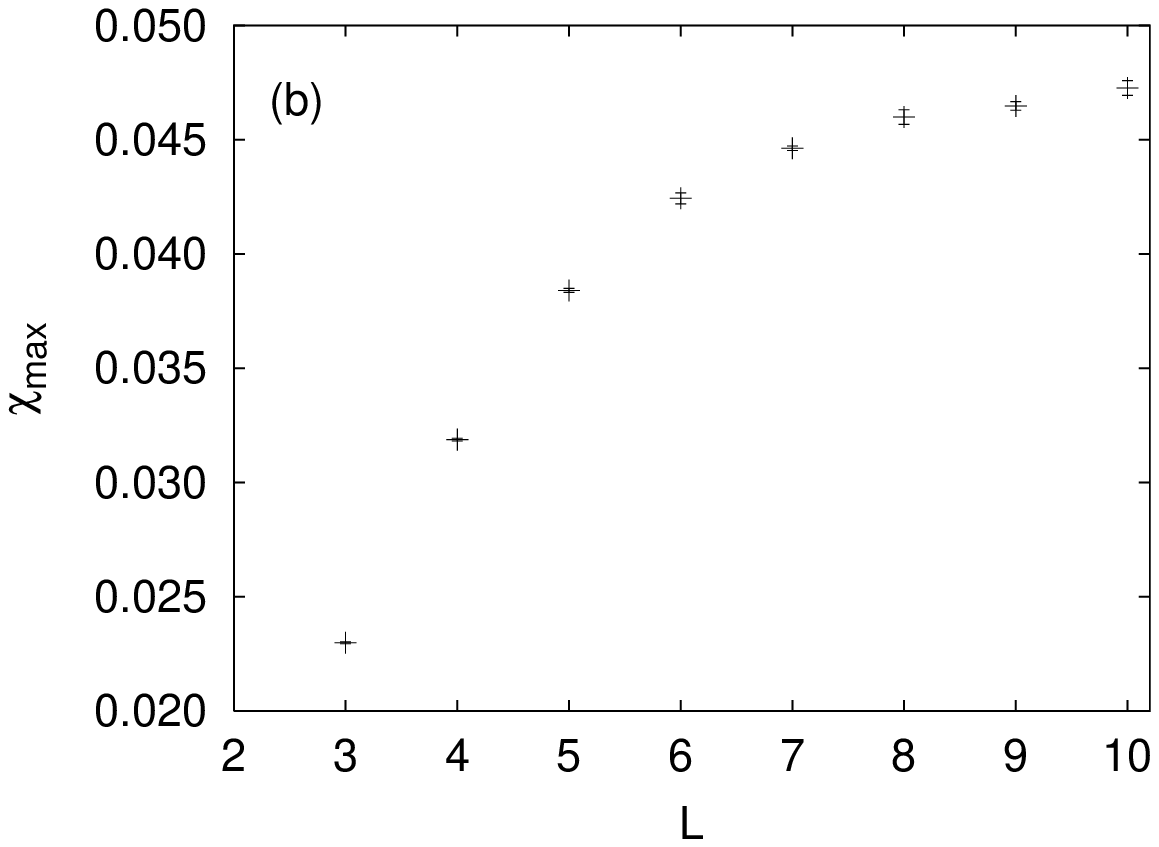,width=8cm,height=6cm}}
\caption{\label{fig1} FSS of (a) the specific-heat maxima $C_{\rm max}$ and (b)
the susceptibility maxima $\chi_{\rm max}$ close to $\beta_c^-$ as a 
function of the lattice size $L$. In particular the behavior of $\chi_{\rm max}$ is 
indicative for a cross-over rather than a true (continuous) phase transition.
}
\end{figure}

To determine the order of the transition at negative $\beta$ we employed
in Ref.~\cite{mel} histogram techniques 
and  used the Binder--Challa--Landau (BCL) cumulant criterion \cite{B1}.
The BCL cumulant is defined as
$B_L := 1- { \langle E^4 \rangle \over 3 \langle E^2 \rangle ^2 }$, 
with $E$ being the action of the system under consideration.
It was evaluated for  $A_t \delta_t$ on different lattice sizes with $L=3$
to $10$ vertices per direction, simulating the system at several values of
the bare coupling $\beta$ with high statistics (using always $\epsilon=0.0875$ and 
$\lambda=0$).
For the BCL cumulant a trend towards 2/3 was observed and all histograms 
showed a clear one-peak structure, cf. Ref.~\cite{mel}.
In our recent Monte Carlo simulations~\cite{a2} we typically generated $200\,000 - 500\,000$
iterations, and recorded for every run the time series of the energy density
$e=E/N_0$ and the magnetization density $m= \sum_l \sigma_l /N_0$, where $N_0=L^4$
is the lattice size.
To obtain results for the various observables ${\cal O}$ at values of the bare 
gravitational
coupling $\beta$ in an interval around the simulation point $\beta_0$, 
the reweighting method~\cite{Ferr} was applied.
By this means we can compute the specific heat,
\begin{equation}
C(\beta)=\beta^2 N_0 (\langle e^2 \rangle- \langle e\rangle^2)~,
\end{equation}
and the (finite lattice) susceptibility,
\begin{equation}
\chi(\beta)=N_0(\langle m^2 \rangle -\langle |m| \rangle^2)~,
\end{equation}
in a certain $\beta$-range around the simulation point.
Figure \ref{fig1} shows the finite-size scaling (FSS) of
the maxima of the specific heat $C_{\rm max}$ and the susceptibility
$\chi_{\rm max}$, respectively. While the behavior of
$C_{\rm max}$ could still be explained as critical scaling with a negative 
exponent $\alpha$, the flattening of $\chi_{\rm max}$ as $L$ increases is quite unusual for
a second-order transition and should rather be taken as an indication for a 
cross-over regime.

\begin{figure}[t]
\centerline{
\psfig{figure=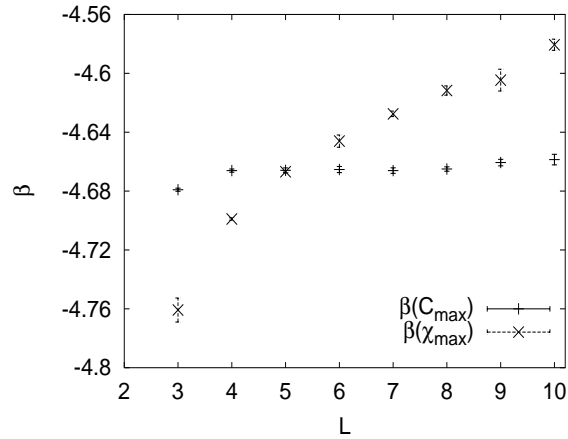,width=8cm,height=6cm}
}
\caption{\label{fig2} Volume dependence of the (pseudo-)critical gravitational couplings
$\beta(C_{\rm max})$ and $\beta(\chi_{\rm max})$, as determined from
the locations of $C_{\rm max}$ and $\chi_{\rm max}$ shown in figure \ref{fig1}.}
\end{figure}

Another feature of the system can be seen in figure~\ref{fig2}
depicting the critical gravitational coupling, as determined from the
specific-heat maxima $C_{\rm max}$ and the susceptibility maxima 
$\chi_{\rm max}$.
In the case of a second-order transition the extrapolations of all 
pseudo-transition points lead to one infinite-volume critical value. 
This seems to be violated in the four-dimensional Discrete Regge model and 
thus again favors the interpretation as a cross-over phenomenon over a true,
thermodynamically defined phase transition.

\section{Summary and conclusion}

We explored the complete phase structure of three different formulations of
lattice quantum gravity in four dimensions including the region of negative
gravitational couplings~\cite{mel}. The qualitative resemblance of the results from simulations of
the $Z_2$-Regge Model to those with continuously varying edge lengths is
particularly remarkable. We computed two-point functions close to the critical
bare gravitational coupling in the negative coupling regime. In the case of
the Group Theoretical Model as well as for the $Z_2$-Regge Model we found 
some evidence for long-range correlations corresponding to massless gravitons.
Eventually, if there is universality between SRC, GTM, and
$Z_2$RM, the same results for physical quantities  are to be expected in 
the continuum limit.

In the present analysis of the scaling of the maxima of the specific heat 
and susceptibility we found evidence for a cross-over regime in the 
four-dimensional Discrete 
Regge model of quantum gravity in the negative coupling region~\cite{a2}.
If this can be substantiated by further investigations and also for the Regge theory with
continuous link lengths, the existence of a continuum limit at negative
bare gravitational coupling is questionable. 
This is of major concern
for a continuum theory of quantum gravity with matter fields~\cite{prd02}.

\section{Acknowledgments}
E.B. and W.J. were supported by the EU-Network HPRN-CT-1999-000161 
``Discrete Random Geometries: From Solid State Physics to Quantum Gravity''.

%\Figures
\newpage

\newpage

\newpage

\newpage

\newpage

\newpage

\newpage

%*************** PLEASE DO NOT REMOVE THIS STAR LINE **************

\end{document}